\def\aa{{A\&A}}
\def\aas{{A\&AS}}
\def\aj{{AJ}}
\def\annrev{{ARA\&A}}
\def\apj{{ApJ}}

\def\mnras{{MNRAS}}
\def\nat{{Nature}}

\documentclass[cup5b]{caps}
\usepackage{graphicx}
\usepackage{amssymb}
\usepackage{ociwsymp3}   %Use this for invited papers.

\HeadText{J. H. van Gorkom}  %Enter author name; if three authors, list all three;
\begin{document}

\pagenumbering{arabic}

%Author names should be in captital letters
\author[]{J. H. van GORKOM\\Columbia University, New York}

%Example for multiple authors:
%
%\author[]{D. C. BACKER$^{1}$, A. H. JAFFE$^{2}$, and A. N. LOMMEN$^{3}$
%\\
%(1) University of California, Berkeley, CA, USA\\
%(2) Imperial College, London, England\\
%(3) Sterrenkundig Instituut ``Anton Pannekoek'', Amsterdam, The Netherlands}
%

\chapter{Interaction of Galaxies with the ICM}

\begin{abstract}

Although speculations of an interaction between galaxies and the ICM 
date back more than thirty years, the impact and importance of a 
possible interaction have long remained elusive. 
In recent years the situation has completely
changed. A wealth of data and detailed hydrodynamical 
simulations have appeared that show the  effects of interactions.
Single dish observations show that cluster galaxies are deficient in
their neutral hydrogen content out to two Abell radii. 
The deficient galaxies tend
to be on radial orbits. Detailed imaging of the neutral
hydrogen distribution in individual galaxies in two nearby clusters show  
a remarkable trend of H~I extent with location in the cluster. 
These trends can be reproduced in simulations of ram pressure stripping
by the ICM using SPH and full 3D hydro-codes. Detailed imaging
of individual galaxies have found  a number of galaxies with undisturbed
stellar disks, truncated gas disks that are much smaller than the stellar
disks, asymmetric extraplanar gas in the center and enhanced
central star formation. These phenomena have all been predicted
by hydrodynamical simulations. For the first time detailed observations
of gas morphology 
and kinematics are used to constrain simulations. Simple models
of ram pressure stripping are consistent with the data for some
galaxies, while for other galaxies more than one mechanism must be at work. 
Optical imaging and spectroscopic surveys show that small H~I disks go
together with truncated star forming disks, that hydrogen deficiency
correlates with suppressed star formation rates and that the spatial extent
of H~I deficiency in clusters is matched by or even surpassed by the extent
of reduced star formation rates. 

Recent volume limited imaging surveys of clusters in the local universe 
show that most gas rich galaxies are located in smaller groups and subclumps, 
that yet have to fall into the clusters. These groups form an ideal
environment for interactions and mergers to occur and we see much evidence
for interactions between gas rich galaxies.

\end{abstract}

\section{Introduction}

It has long been known that in the local universe the mix of morphological 
types differs in different galactic environments with ellipticals and 
S0's dominating in the densest clusters 
and spirals dominating the field population (Hubble and Humason 1931). 
This so called
density-morphology relation has been quantified by Oemler (1974) and  Dressler
(1980) and is found to extend over five orders of magnitude in space
density (Postman and Geller 1984). Whether this relation arises at formation
(nature) or is caused by density driven evolutionary effects (nurture)
remains a matter of debate. More recent studies of clusters of
galaxies at intermediate redshifts show that both the morphological mix 
and the star formation rate strongly evolve with redshift (Poggianti et
al. 1999; Dressler et al. 1997; Fasano et al. 2000). In particular the
fraction of S0's goes down and the spiral fraction and star formation
rate go up with increasing redshift. 
There are many physical mechanisms at work in clusters or  
during the growth of clusters that could affect the star formation rate
and possibly transform spiral galaxies into S0's. In this review
I will limit myself to the role that the hot intracluster medium (ICM)
may play. 
 
The first suggestion that an interaction between the ICM and disk galaxies
may affect the evolution of these galaxies was made immediately after
the first detection of an ICM in clusters (Gursky et al. 1971). In a seminal 
paper  on ``the infall of matter into clusters'' 
Gunn and Gott (1972) discuss what might happen
if there is any intergalactic gas left after the clusters has collapsed. 
The interstellar
material in a galaxy would feel the ram pressure of the intracluster medium
as it moves through the cluster. A simple estimate of the effect assumes
that the outer disk gas gets stripped off when the local restoring
force in the disk is smaller than the ram pressure. Thus disks
gets stripped up to the so called stripping radius where the forces
balance. 
They estimate that for a galaxy moving at the typical velocity of 1700 km/s 
through
the Coma cluster the ISM would be stripped in one pass. This would explain
why so few normal spirals are seen in nearby clusters. In particular 
it would explain the existence of so many gas poor, non star forming disk 
galaxies first noticed by Spitzer and Baade (1951) and later dubbed anemics by
van den Bergh (1976).

Ram pressure stripping is but one way in which the ICM may affect the ISM. 
The effects of viscosity, thermal conduction and turbulence on
the flow of hot gas past a galaxy were considered by Nulsen (1982), who
concluded that turbulent viscous stripping will be an important
mechanism for gas loss from cluster galaxies.  
While the above mentioned mechanisms would work to remove gas from
galaxies and thus slow down their evolution, an alternative possiblity
is that an interaction with the ICM compresses the ISM and leads
to ram pressure induced star formation (Dressler and Gunn, 1983; Gavazzi et al.
1995).

On the observational side there has long been evidence that 
spiral galaxies in clusters have less neutral atomic hydrogen than galaxies
of the same morphological type in the field (for a review
see Haynes, Giovanelli and Chincarini 1984). The CO content however
does not seem to depend on environment (Stark et al. 1986; Kenney and 
Young 1989). 
Both single dish observations and synthesis imaging results of the Virgo 
cluster show that the HI disks of galaxies in projection close to the cluster
center are much smaller than the H~I disks of galaxies in the outer parts
(Giovanelli and Haynes, 1983; Warmels 1988a,b,c; Cayatte et al. 1990, 1994).
 All of these phenomena could easily be interpreted in terms of 
ram pressure stripping. Dressler (1986) made this even more plausible 
by pointing out that the gas deficient galaxies seem statistically 
to be mostly on radial orbits which would carry them into the dense
environment of the cluster core. However nature turned out to be more
complicated than that. In a comprehensive analysis of HI data on six
nearby clusters Magri et al. (1988) conclude that the data can not
be used to distinguish between inbred and evolutionary gas deficiency 
mechanisms or among different environmental effects.  Although H~I deficiency
varies with projected radius from the cluster center,
with the most H~I poor objects
close to the cluster centers, no correlation is found between deficiency
and (relative radial velocity)$^2$, as would be expected from ram pressure
stripping. 
 
In more recent
years a number of developments have taken place. First there was a flurry 
of activity on the theoretical front, for the first time detailed numerical
simulations on the effects of ram pressure stripping appeared. Since then
both improved statistics on HI deficiency and detailed multiwavelengths
observations of cluster galaxies undergoing trauma appeared. More recently
detailed comparisons have been made between individual systems and numerical
simulations. Finally synthesis imaging of neutral hydrogen no longer needs
to be limited to a few selected systems in nearby clusters and results of 
volume limited surveys of entire clusters at redshifts between 0 and 0.2 
have started to appear in the literature.
In this review I will first discuss what we have learned about the statistical
properties of the H~I content of cluster galaxies. Then I will review
some of the recent numerical work that has been done and compare these with 
observational results. After that I will discuss what we have learned 
from imaging surveys, and in conclusion I will discuss the importance
of the ICM interaction for galaxy evolution. 

  \begin{figure}
    \centering
    \includegraphics[width=8cm,angle=0]{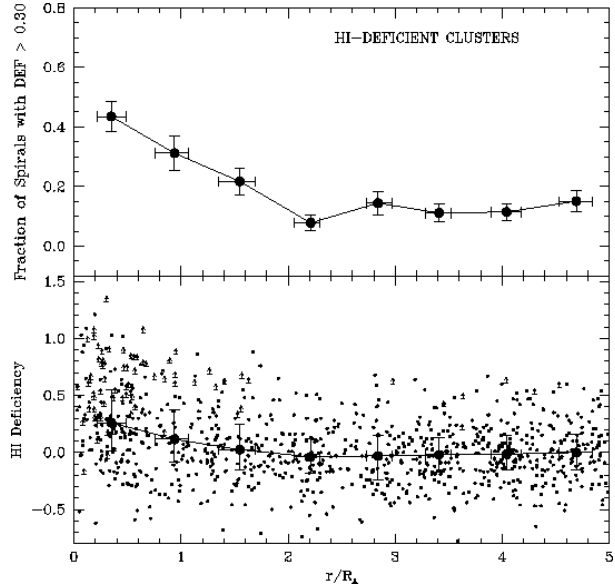}
    \caption{Top: H~I deficient fraction in bins of projected radius 
from the cluster center for the superposition of all the H~I deficient 
clusters. Bottom: H~I deficiency versus projected radius from the cluster 
center. Small dots show   the radial variation of H~I deficiency for 
individual galaxies, while the arrows indentify non detections plotted
at their estimated lower limit. Large dots are the medians of the 
binned number distribution. From Solanes et al. 2001.}
    \label{}
  \end{figure}

\section{The Statistics of H~I Deficiency}

The most comprehensive survey on H~I content in cluster galaxies
to date is the work by Solanes et al. 2001. These authors compiled H~I data on
1900 spiral galaxies in 18 nearby clusters. The data are mostly
obtained with the Arecibo telescope, a single pixel telescope, and
give information about the total amount of neutral hydrogen within
the Arecibo beam (3 arcmin) centered on optically selected galaxies. 
Galaxies are earmarked as belonging to a cluster when they fall within
a projected distance of 5 Abell radii (R$_A$), i.e. within 7.5 h $^{-1}$ Mpc,
from the cluster center and  have a radial velocity that is less
than 3 times the average velocity dispersion from the cluster mean. 
Only clusters are included, for which there are good H~I data for at
least ten galaxies within 1 R$_A$ of the cluster center. HI defiency
is calculated according to the recipe of Haynes and Giovanelli (1984).
It is the log$_{10}$( M$_{H~I}$ observed/ M$_{H~I}$ expected), where the
expected H~I mass is derived from a sample of isolated spirals of the 
same morphological type and optical diameter.
To get significant statistics the cluster sample is then divided
in two groups, the deficient cluster sample, and the non deficient
cluster sample. The deficient sample contains all clusters for 
which the H~I deficiency distribution over galaxies is significantly 
different within 1 R$_A$ from that of the galaxies outside 1 R$_A$. 
One of the most remarkable results of the analysis of that data base
is shown in Figure 1.  It shows the H~I deficient fraction, i.e.
the fraction of galaxies with a deficiency greater than 0.3, 
in bins of projected radius from the cluster center for the superposition
of all the H~I deficient clusters.
Galaxies with a deficiency greater than 0.3,
are galaxies that are deficient in neutral hydrogen by a factor two or
more as compared to isolated galaxies of the same morphological type and size 
in the field.  
The percentage  of H~I deficient spirals increases
monotonically going inward. What is surprising is that this monotonic rise
starts as far out as two R$_A$. This suggests that the effect of the cluster
environment can be felt out to two Abell radii, far beyond the reaches of the 
dense ICM. 

  \begin{figure}
    \centering
    \includegraphics[width=10cm,angle=0]{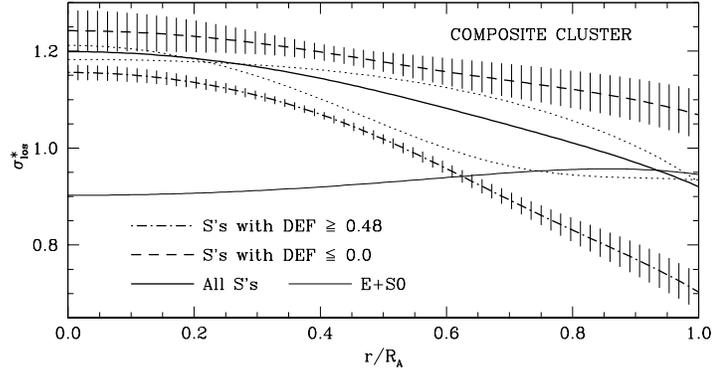}
    \caption{ Radial run of normalized line of sight velocity dispersion
for the composite H~I deficient cluster. From Solanes et al. 2001.}
    \label{}
  \end{figure}

No correlation is found for the fraction of H~I deficient
spirals (i.e. the number of spirals with an H~I deficiency DEF $\ge$ 0.30
within 1 R$_A$ of the cluster center compared to all galaxies
of that type found in that region) with global cluster properties,
such as X-ray luminosity, X-ray temperature, and radial velocity dispersion.
As pointed out in the paper, this could be a selection bias. 
If stripped spirals would lose all their gas they may be transformed into
S0's and they would be left out from the statistics. It is  
somewhat plausible that this is indeed the case since the fraction of spirals
is clearly anti correlated with X-ray luminosity.
The most important result of the paper apart from the extent of
the occurrence of deficient galaxies is the correlation between 
deficiency and orbital parameters. This is shown in Figure 2.
It shows for the composite deficient cluster the radial run of the line of 
sight velocity dispersion for the most deficient spirals, the non deficient 
spirals,
for all spirals and for ellipticals and lenticulars.
If galaxies are on radial orbits the measured velocity dispersion should
decrease at large distances from the cluster center. Although all spirals
show a decrease in velocity dispersion at large distances, this effect 
is by far the most pronounced in the H~I deficient spirals. The ellipticals
and S0's have a constant velocity dispersion with radius. This confirms
the result by Dressler (1986). These results suggest that deficiency
is most pronounced when galaxies go through the dense cluster center at 
high velocities and as such support the idea that ram pressure
stripping causes the deficiency.

\section {Simulations}

In the early seventies several papers appeared considering the 
effect of the ICM on galaxies in clusters, e.g. ram pressure
stripping (Gunn and Gott 1972), evaporation (Cowie and Songaila 1977)
and turbulent viscosity and evaporation (Nulsen 1982). Not much
work was done though to connect theory with observations. 
The first paper that specifically looks at observational 
characteristics is the paper by Stevens, Acreman and Ponman (1999).
This paper focusses on the impact of the ICM on an elliptical galaxy
with a hot ISM and calculates the observational signatures of this
in the hot ICM, predicting bow-shocks, wakes and tails. 
Some, but still precious few, examples of structures that could be 
interpreted like this exist in the X-ray literature (Stevens et al. 1999).
Observationally, there is much more evidence for the impact
of the ICM on the cool ISM of disk galaxies. From a view point of
galaxy evolution this is also the more important question. In clusters
star formation rates are known to evolve rapidly between intermediate 
redshifts and the local universe (Poggianti et al 1999, Balogh et al 1999).
A mechanism is required to bring star formation almost completely
to a halt and a major issue is whether ram pressure stripping could 
do this to disk galaxies. The original analytical estimates of Gunn and 
Gott (1972) predict that gas gets stripped from a galaxy up to 
a stripping radius within which the restoring force from the disk 
exceeds the ram pressure. The first numerical simulations 
(Abadi, Moore and Bower (1999)) using a 3 dimensional SPH/N-body
simulation to study ram pressure stripping of gas from spiral galaxies
orbiting in clusters confirm that gas 
in disk galaxies gets stripped up to the stripping radius estimated
by Gunn and Gott (1972). At small radii the potential provided by the
bulge component contributes considerably. They estimate that a galaxy
passing through the center of Coma would have its gaseous
disk truncated to $\approx$ 4 kpc, losing about 80 $\%$ of its gas.
However the process is in general not efficient enough to account for the
rapid and widespread truncation of star formation observed in cluster
galaxies. 
Quilis, Moore and Bower (2000) use a finite difference code 
to achieve higher resolution in order to be able to include complex
turbulent and viscous stripping at the interface of cold and hot
gaseous components as well as the formation of bow shocks in the ICM ahead
of the galaxy. From only a few selected runs on galaxies with holes
in the central gas distribution they reverse the conclusion of Abadi
et al. (1999) and state that ICM - ISM interaction could explain
the morphology of S0 galaxies and the rapid truncation of star formation
implied by spectroscopic observations. The main difference with the
Abadi et al. result is the use of a complex multi phase structure of the ISM.
They show that the presence of holes and bubbles
in the diffuse H~I can greatly enhance the stripping efficiency.
As the ICM streams through the holes in the ISM it ablates the edges and 
prevents stripped gas from falling back. 
Schulz and Struck (2001) in a comprehensive study using SPH,
an adaptive mesh HYDRA code, and including
radiative cooling, confirm that low column density gas
is promptly removed from the disk. They also find that the onset of the
ICM wind has a profound effect on the gas in the disk, that does 
not get stripped. The remnant disk is compressed and slightly displaced
relative to the halo center. This can trigger gravitational instability,
angular momentum gets transported outward and the disk compresses further
forming a ring. This makes the inner disk resistant to further
stripping, but presumably susceptible to global starbursts.
These various simulations appear to more or less agree on the effects 
of the ISM. All of the above work modelled the ICM as a constant wind. 
Vollmer et al. (2001) took a different approach. Using an N-body/sticky-particle
code they simulate galaxies in radial orbits through the gravitational 
potential of the Virgo cluster. The galaxies thus experience a time variable
ram pressure and  maximal damage to their gaseous disks only becomes apparent
well after closest approach to the Virgo Cluster
center. Thus if we see galaxies with truncated H~I disks or distorted
velocity fields they are likely to be on their way out from the center.
They also find that a considerable part of the stripped total gas mass 
remains bound to the galaxy and falls back onto the galactic disk 
after the stripping
event,  possibly causing a central starburst. The results of Schulz and
Struck (2001) and Vollmer et al (2001) are the first to produce
simultaneously stripping in the outer parts and a mechanism to
enhance star formation in the inner parts. This may help use up any
remaining gas in the central regions and it could possibly do some
secular bulge building. There is observational evidence for stripped
H~I disks with enhanced central H~I surface densities (Cayatte et al. 1994)
and there are several lines of evidence that the most recent episode
of star formation in cluster galaxies occurred in the central parts 
(e.g. Rose et al. 2001)

  \begin{figure}
    \centering
    \includegraphics[width=10cm,angle=0] {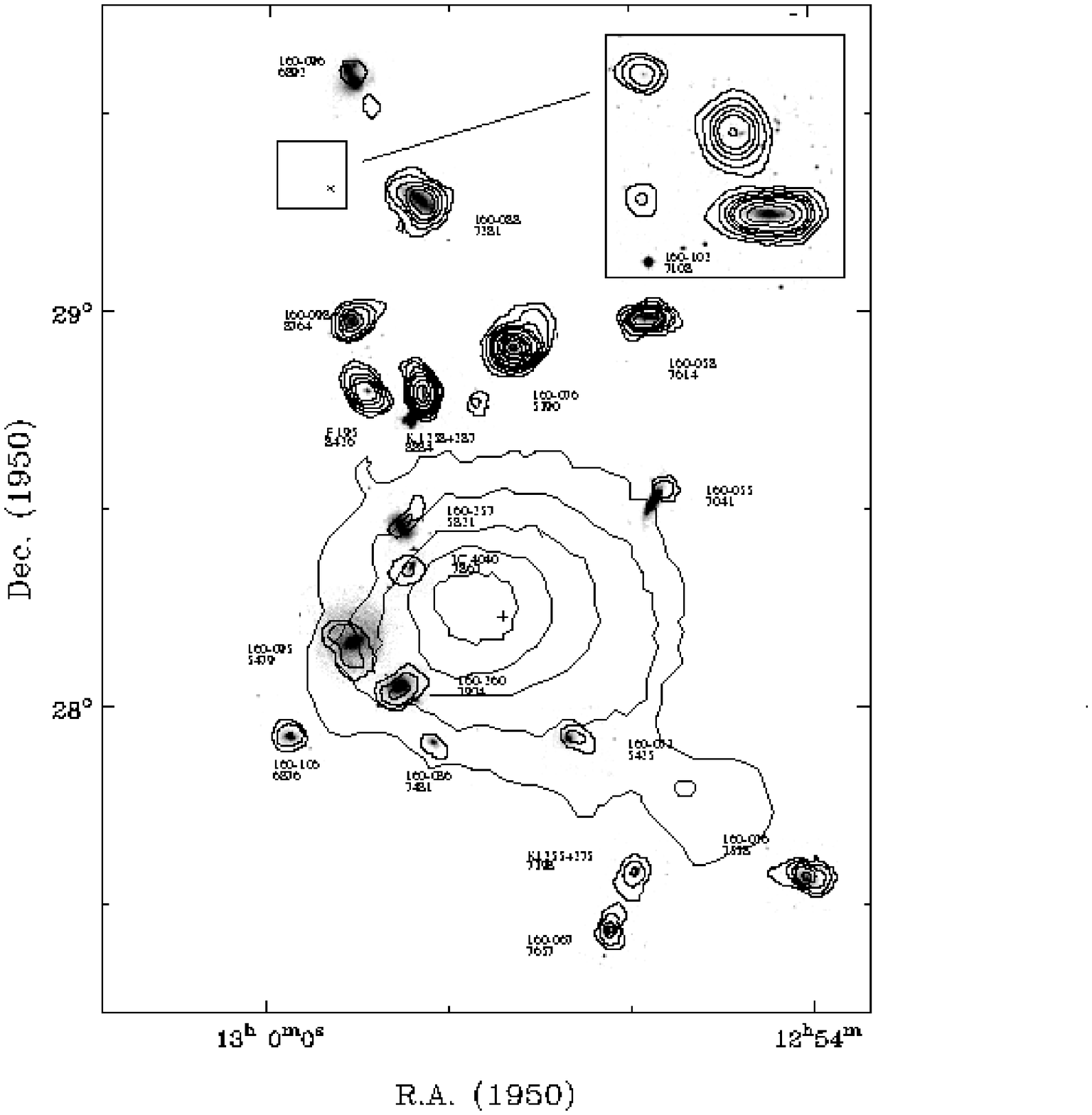}
    \caption{Composite of individual H~I images of Coma spirals observed
with the VLA. Galaxies are shown at their proper position and they 
are magnified by a factor 7. The H~I images (contours) are overlaid 
on DSS optical images (greyscale). The large scale contours sketch the
X-ray emission as observed by Vikhlinin et al. (1997).
From Bravo-Alfaro et al. 2000.}
    \label{}
  \end{figure}

  \begin{figure}
    \centering
    \includegraphics[width=10cm,angle=0]{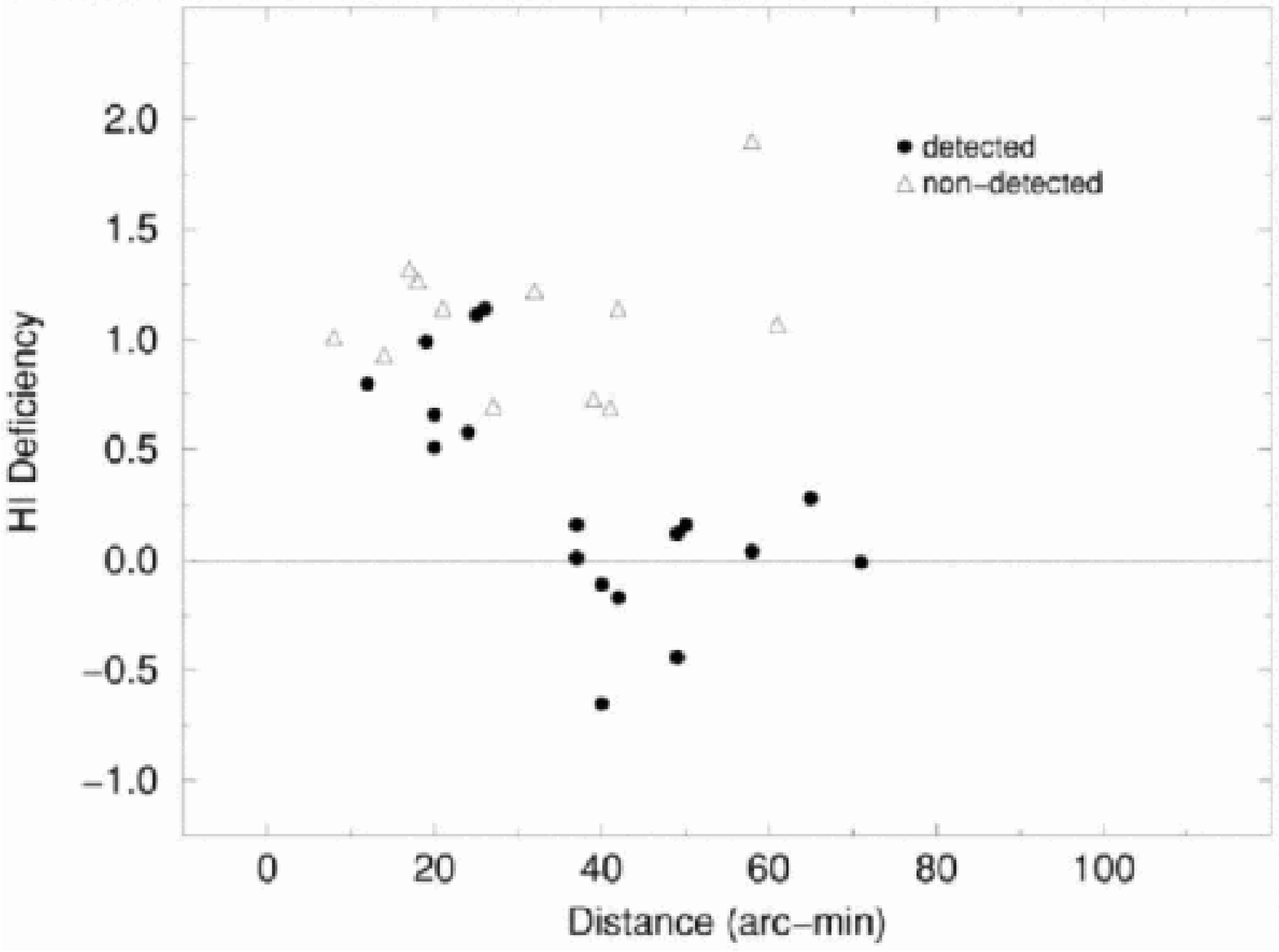}
    \caption{Distribution of H~I deficiency parameter 
as a function of the projected distance from the center of Coma.
Filled circles correspond to H~I detected galaxies and triangles
to the lower limits of the deficiency parameter for the galaxies
that are not detected in H~I. From Bravo-Alfaro et al. 2000.}
    \label{}
  \end{figure}

\section{Comparison of Simulations with H~I Imaging}

The wealth of single dish data on the
H~I content of selected spirals in the cluster environment,
has shown beyond doubt that H~I deficiency occurs among cluster 
spirals. These data are less suitable to study the mechanisms that
remove the gas. Projection effects along the line of sight and
uncertainties about the orbital history of individual galaxies complicate 
matters. 
H~I imaging has so far provided far
less statistics, but in the imaging data individual galaxies can
be selected that appear to have distortions in their H~I morphology
or kinematics that are unique to the cluster environment.  
A prime example is the occurrence of
tiny H~I disks in Virgo (Cayatte et al. 1990; Warmels 1988a,b).
The size of the gaseous disks is considerably smaller than the 
optical disk, the effect is most pronounced close to the cluster center
and gently decreases at increasing distance from the center.
In addition to Virgo this has now also been seen in the Coma cluster 
(Bravo-Alfaro et al. 2000, 2001). Figure 3 shows an overlay of the total 
H~I emission (contours) on an DSS optical image in greyscale. Each galaxy
is located at its proper position in Coma, but the images are blown 
up by a factor 7. The thick contours are the X-ray emission as observed
with  ROSAT. The first thing to note is that the H~I disks seen in projection
on the X-ray emission are in general smaller compared to the optical 
image than for galaxies far from the center of Coma. An example in case
is the galaxy CGCG 160-095 (NGC 4921) east of the center where H~I
is only seen in one half of the disk. This must be caused by a mechanism that 
only affects the gas and ram pressure stripping is a good candidate.
 Figure 4 shows the H~I deficiency
versus projected distance from the center. Its interpretation is already more
complicated. Though none of the galaxies projected on to the X-ray emission
has a normal H~I content (deficiency 0.0), there are several
non detections out to large projected radii. Possibly these galaxies
have already gone through the center.

  \begin{figure}
    \centering
    \includegraphics[width=8cm,angle=0]{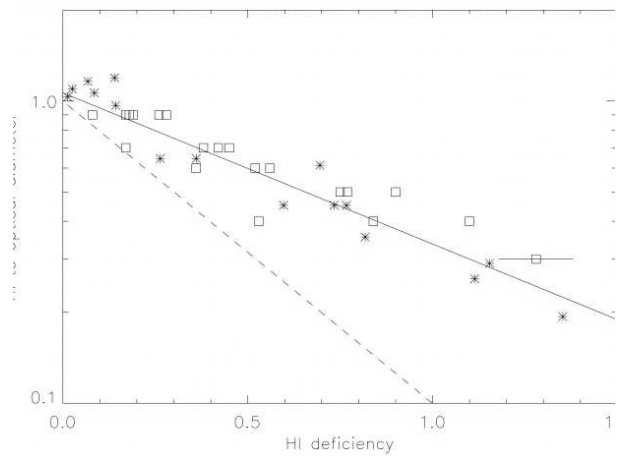}
    \caption{Normalized H~I to optical diameter as a function
of the H~I deficiency for the Virgo Cluster. 
Squares: observed values (Cayatte et al. 1994);
stars: model values. The solid line assumes that the H~I surface density
has the same value before and after the stripping event.
From Vollmer et al. 2001}
    \label{}
  \end{figure}

Imaging of Virgo and Coma indicates
that H~I disks that are smaller than the optical ones may be generic
to cluster galaxies. Numerical simulations show that this is what
you expect from ram pressure stripping. Abadi et al. (1999) show 
the results of a simulation with a constant ram pressure typical of the
Virgo Cluster ICM and galaxies moving with relative velocities 
of 1000 km/s for galaxies of different size. The dependence of 
stripping radius on disk scale length (plotted in their Figure 3) 
is consistent with the result found by Cayatte et al. (1994).
Even more impressive is the result
of Vollmer et al. (2001) shown in Figure 5. The simulation specifically
models galaxies on radial orbits through the Virgo potential. The figure
plots H~I deficiency versus H~I to optical diameter. Both the model
data (stars) and observed values (Cayatte et al. 1994) are shown. The solid 
line corresponds to a model where only the outer parts of the disk get 
stripped and the constant central H~I radial surface density remains 
unchanged in the stripping process. 
 
Vollmer in a series of papers (Vollmer 2003; Vollmer et al. 2001b; 
Vollmer et al. 2000; Vollmer et al. 1999) tries to reproduce observed
gas distribution and kinematics of selected Virgo spirals with
his N-body/sticky particle simulations. The galaxies are selected 
based on their
H~I morphology and all show truncated H~I disks. This is the first time
ever that both gas distribution and kinematics are put to test in comparison
with models. In several cases the simulations can reproduce the observed 
signatures in the gas and all of these galaxies are found to be on their
way out of the cluster having passed through the dense center. 
In  one case (NGC 4654) Vollmer (2003) shows
that both ram pressure stripping and a gravitational interaction must
be at play.

  \begin{figure}
    \centering
    \includegraphics[width=8cm,angle=0]{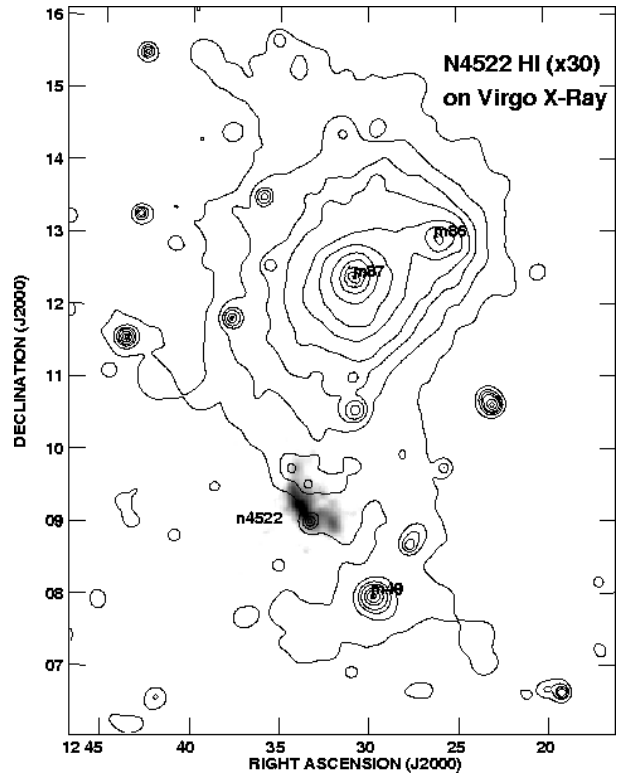}
    \caption{H~I greyscale image of NGC 4522, scaled up in size by a factor
of 30, on a ROSAT X-ray image of the Virgo cluster from 
Bohringer et al. (1994). The image indicates the locations of the giant 
ellipticals M87, M86 and M49, which are associated with sub-clusters.
From Kenney et al. 2003.}
    \label{}
  \end{figure}

  \begin{figure}
    \centering
    \includegraphics[width=10cm,angle=0]{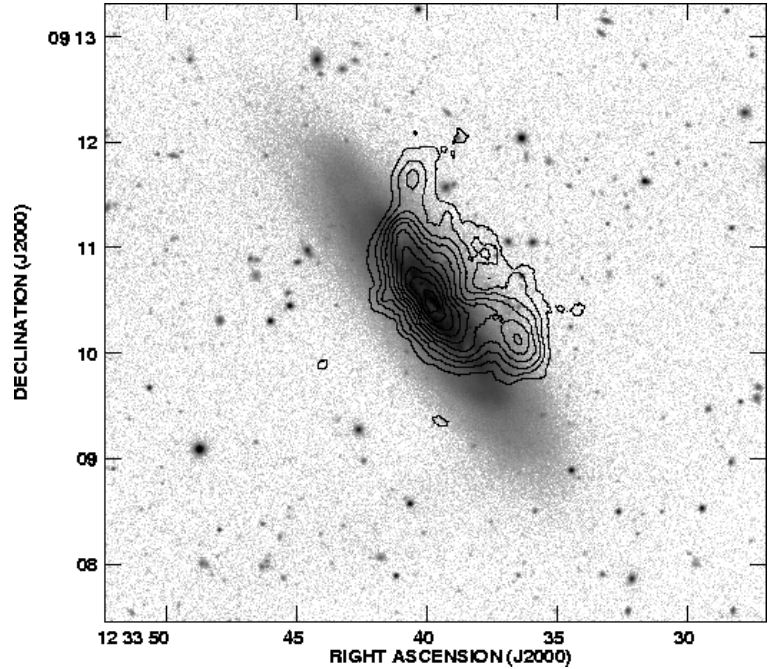}
    \caption{H~I contours overlaid on an R band greyscale image
from the WIYN telescope from Kenney \& Koopmann (1999).
Note the undisturbed outer disk.
From Kenney et al. 2003.}
    \label{}
  \end{figure}

  \begin{figure}
    \centering
    \includegraphics[width=10cm,angle=0]{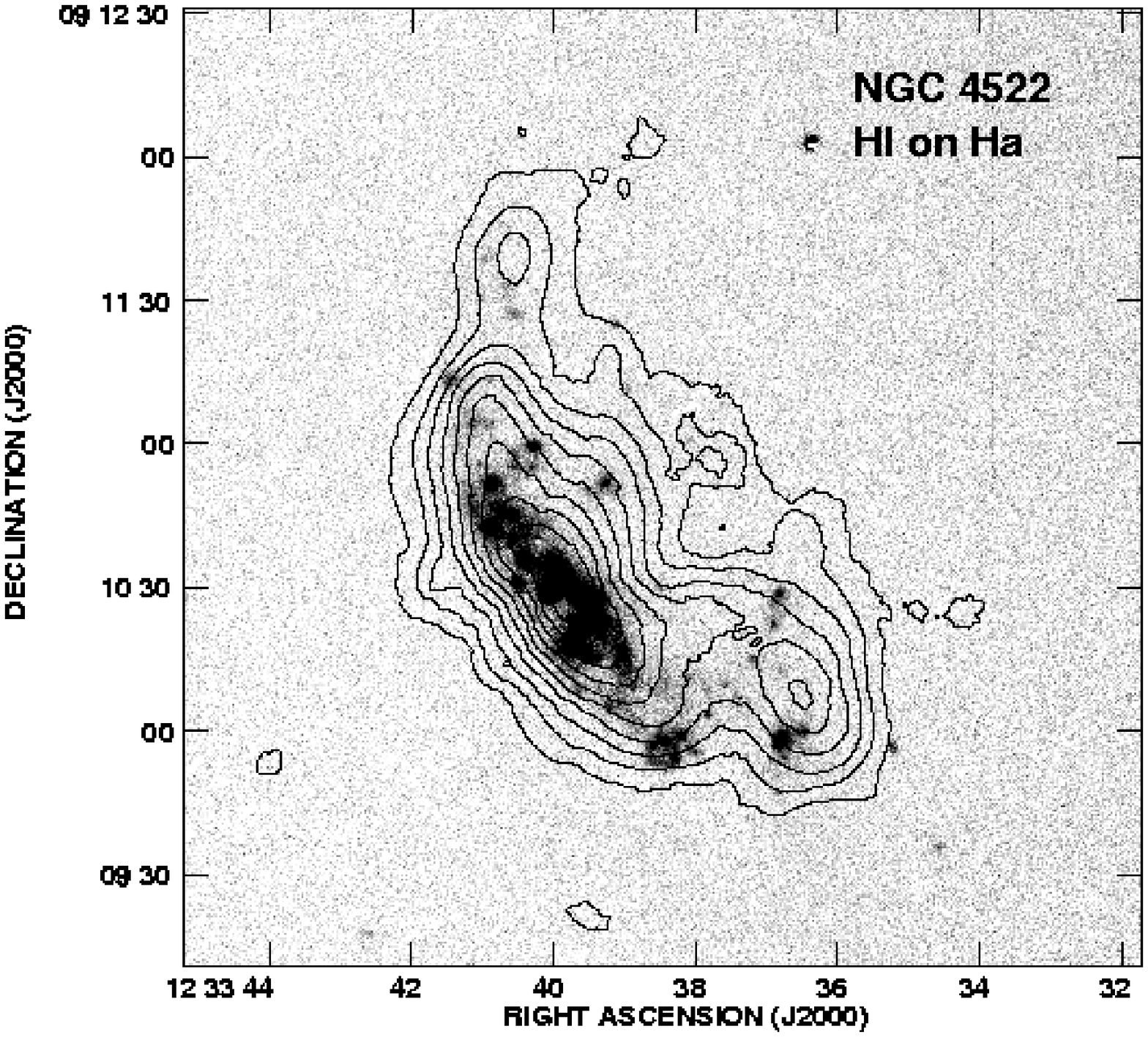}
    \caption{H~I contours overlaid on an H$\alpha$ greyscale
image from Kenney and Koopmann (1999).
Note that there are H~II regions associated with each of the 2 
major extraplanar H~I peaks. 
From Kenney et al. 2003.}
    \label{}
  \end{figure}

One of the most interesting things that has been found recently is a number
of galaxies with truncated H~I disks, normal or enhanced star formation in the
central regions, and some extraplanar gas on one side of the
galaxy, while the stellar disks are completely undisturbed. These may be 
the best candidates for galaxies,
that are currently undergoing an ICM-ISM interaction. 
A prime  example is NGC~4522, studied by Kenney
and collaborators in great detail. Kenney and Koopmann (1999) first
pointed out that NGC~4522  is one of the best candidates for ICM-ISM 
stripping in action. Figures 6, 7 and 8 summarize the main characteristics
of this galaxy. NGC~4522 is located within a subclump of the Virgo
cluster centered on M49. A ROSAT map (Figure 6)  shows
weak extended X ray emission at the projected location of NGC~4522. 
All the known peculiarities of NGC~4522 are associated with gas, dust
and H~II regions, not with the older stars. The H~I (Figure 7) is spatially 
coincident with the undisturbed stellar disk in the central 3 kpc 
(0.4 R$_{25}$)
of the galaxy. At 0.4 R$_{25}$ the H~I truncates abruptly and is 
only seen above the plane to the SW. About half of the total the H~I
appears to be extraplanar extending to $\approx$ 3 kpc above
the plane (Kenney et al. 2003).  There is striking similarity between the 
spatial distribution  of the H$\alpha$ emission and the H~I emission 
(Figure 8). The H${\alpha}$ emission from the disk is confined to the inner 
3 kpc as well and  extraplanar H$\alpha$ filaments (10$\%$ of the H$\alpha$
emission) emerge from the outer edge of the H$\alpha$ disk (Kenney and
Koopmann 1999). Note that there are H~II regions associated with each of 
the 2 major extraplanar H~I peaks, and that those in the SW are much
more luminous. These data strongly suggest an ICM-ISM interaction.
The undisturbed stellar disk rules out a gravitational interaction.
The truncated H~I disk suggests ram pressure stripping is at work.
The extraplanar gas  has almost certainly been swept out of the disk.
A detailed simulation by Vollmer et al. 2000 suggests that the gas is falling
back after a stripping event, but this is inconsistent with the
observed kinematics (Kenney et al. 2003). More likely the gas is still on 
its way out due to ram pressure stripping (Kenney and Koopmann 1999).
The combined characteristics such as a stripped H~I disk, only central
H$\alpha$ emission in the disk, extraplanar gas on only one side of the
galaxy make this such a convincing candidate for an ICM-ISM interaction.
Evidence for enhanced central star formation in some other examples like this
makes this the more interesting.
A central starburst would help use up any remaining gas and a
morphological transformation would be in place.

  \begin{figure}
    \centering
    \includegraphics[width=10cm,angle=0]{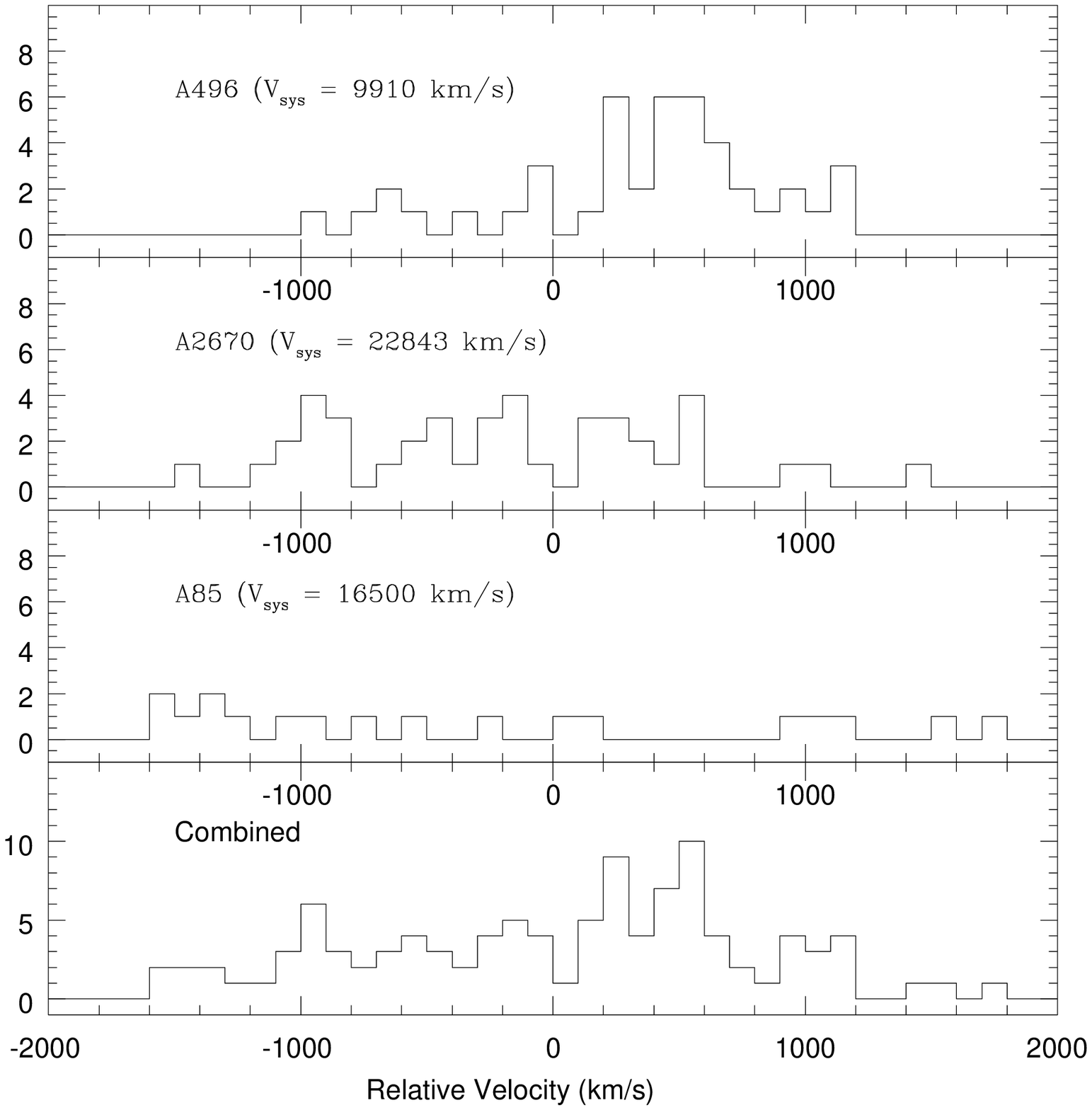}
    \caption{The velocity distribution of the H~I detected galaxies
in Abell 496, 2670 and 85 with respect to the mean velocity of the 
cluster. At the bottom the combined distribution for the three clusters.
Note the very non gaussian distribution of the velocities. In Figure
11 an example is shown of spatial and velocity clustering of
H~I detected galaxies.}
    \label{}
  \end{figure}

  \begin{figure}
    \centering
    \includegraphics[width=10cm,angle=-90.0]{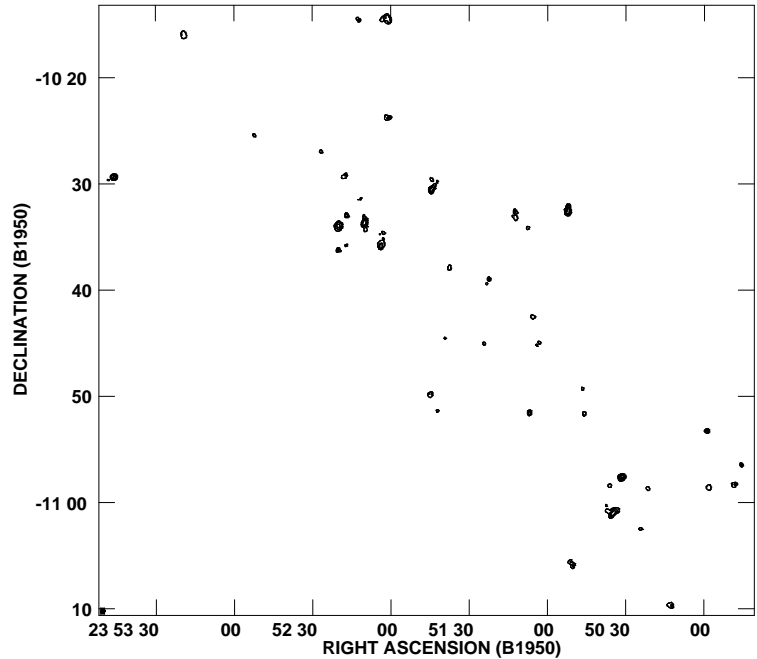}
    \caption{Total H~I emission of the cluster Abell 2670
 at a redshift of z=0.08. The image
is centered on the center of the cluster and the H~I detections 
are spread over a region of 5x2 h$^{-1}$ Mpc. All galaxies with an H~I mass
 $\ge$ $2 \times 10^8$ M$_{\odot}$ in the velocity range 
of the cluster are shown.}
    \label{}
  \end{figure}

  \begin{figure}
    \centering
    \includegraphics[width=10cm,angle=0]{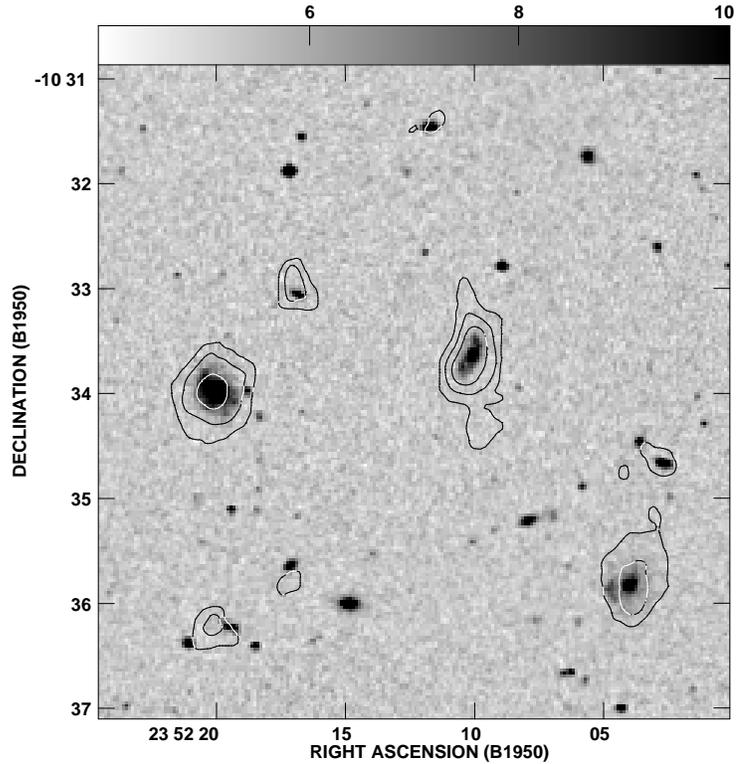}
    \caption{An overlay of the total H~I emission (contours)on an optical
image of the DSS (greyscale) of a group of galaxies in the NE part
of the A2670 cluster. The internal velocity dispersion of this 
group is only a few 100 km/s. These are very gas rich galaxies.
Note the distorted H~I contours indicative of interactions.} 
    \label{}
  \end{figure}

  \begin{figure}
    \centering
    \caption{Recently discovered H~I filaments near the 
SO galaxy, NGC 4111, in the Ursa Major cluster. The SO galaxy is located 
in a small group and the H~I morphology indicates that tidal interaction
between the galaxies is taking place. From Verheijen and Zwaan (2001).}
    \label{}
  \end{figure}

\section{Surveys and the Importance of Interactions with the ICM}

So far I have only presented evidence that interaction with the
ICM occurs based on selected cases. The small gaseous disks 
in the optically selected samples of Virgo and Coma are almost
certain due to rampressure stripping. A growing number of spiral galaxies is
found with an unusual morphology in HI, H$\alpha$, and
radio continuum, e.g.  the long known examples in A1367 (Gavazzi et al. 1995),
in Virgo galaxies such as NGC 4522, NGC 4388 (Veilleux et al. 1999), 
NGC 4569 and NGC 4438
(Vollmer, in prep) and in Coma (Bravo-Alfaro et al. 2001, Gregg 2003,
Beijersbergen 2003). These are prime candidates for
ongoing ram pressure stripping. 

How important are these ICM-ISM interactions for the evolution of galaxies
in clusters? An important first step to address this question is the imaging
study by Koopmann and Kenney (1998, 2002) of 55 Virgo Cluster spirals
in H$\alpha$ and R band. They find that the total massive star formation
rates in Virgo Cluster spirals have been reduced by factors up to 2.5
in the median compared to isolated spirals. The reduction in total star 
formation is caused primarily by truncation of the star-forming disks
(seen in 52$\%$ of the spirals). Some of these have undisturbed stellar
disks and are likely the product of ICM-ISM stripping, but others
have disturbed stellar disks, and are likely the product of tidal
interactions or minor mergers, possibly in addition to ICM-ISM stripping.
Some evidence is found for enhanced star formation rates due to
low velocity tidal interactions and possibly accretion of H~I gas.
A strong correlation is found between H~I deficiency and normalized
H$\alpha$ flux. The authors conclude that the survey provides strong
evidence that ICM-ISM interactions play a significant role in the 
evolution of most Virgo spirals by stripping gas from their outer disks. 

So far I have only discussed H~I results obtained on individual galaxies that
were selected because they were interesting or, at best, because they
were in some optical flux limited sample. To see how the gas content and 
morphology 
depends on cluster environment  optically unbiased studies need to be done. 
Ideally one should
probe the entire volume of clusters, including the low density outskirts,
to get some idea of the gas content and star formation properties as
function of local or global density. 

The first volume limited H~I survey of a cluster was done of the Hydra cluster 
(McMahon 1993; van Gorkom 1996). Dickey (1997) surveyed two clusters in the 
rich group of clusters in the Hercules cluster. These surveys already
show that there is a great variety in the neutral hydrogen properties
of clusters. Hydra shows barely any evidence for hydrogen
deficiency, despite the fact that it is very similar in its global
properties to the Virgo Cluster.  The most likely explanation of these
results is that Hydra is in fact a superposition of at least three
groups along the line of sight, seen in projection close to each other.
The most striking result of the Hercules survey (Dickey 1997) is 
the spatial variation of H~I properties within the clusters.
Galaxies in the A2147 and 
the southwest of A2151 show strong H~I deficiency, while galaxies
in the northeast of A2151 are gas rich. It is perhaps one of the
most convincing demonstrations of environmental impact on galaxy properties.
The X-ray luminous clusters have strong H~I deficiency, the parts
of the clusters that have no detectable ICM have an abundance of gas 
rich galaxies. 

A more systematic survey of five nearby clusters 
(Abell 85, 754, 496, 2192, 2670) is currently being done at the VLA
(van Gorkom et al. 2003; Poggianti and van Gorkom 2001).
Each cluster is completely covered out to 2 R$_A$ thus covering the
dense inner parts and the low density outer parts and the entire
optical velocity range is probed. The most striking result is
that in all clusters the H~I detections are highly clustered both spatially
and in velocity. Figure 9 shows the velocity distribution of the H~I 
detections in three of the clusters. Although the velocity distibution
of the optically catalogued galaxies in each of the clusters is gaussian,
the velocity distribution of the gas rich galaxies is far from gaussian.
Figure 10 shows the total H~I image of Abell 2670. Contours 
represent the integrated H~I emission for individual galaxies. 
At first glance the image looks like the images of the Virgo
and Coma cluster with small H~I disks close to the center and large H~I disks 
further out. But this image now shows the H~I emission from all galaxies 
in the cluster with H~I masses $\ge$ $2 \times 10^8$ M$_\odot$. Figure 11
shows an overlay of a group of galaxies to the NW on the DSS. The internal
velocity dispersion of this group is only a few 100 km/s. These galaxies
are very gas rich and obviously conditions for interactions and merging
are ideal. Several galaxies do in fact show evidence for distorted H~I. 
These results indicate that gas rich 
disk galaxies that make it into the center of a cluster are likely to be 
seriously affected by interaction with te ICM. It is likely that a
significant fraction of disk galaxies, falling into clusters, is
located in low velocity dispersion loose groups. The interactions 
in these groups, before
the actual infall, may be more damaging to the morphology than any ICM
interaction thereafter. The most dramatic example (Figure 12) of that is the 
H~I image of a number of S0 galaxies in the outskirts of the
Ursa Major cluster by Verheijen and Zwaan (2001). Optically one would not
have guessed that anything dramatic is about to happen to these galaxies.
The H~I shows that strong interactions are already taking place.

\section{Concluding remarks}

We can now begin to answer the question posed in the introduction:
are interactions with the ICM important for the evolution of disk galaxies.
The answer is a definite yes for disk galaxies in cluster environments.
We see individual galaxies that show all the signs of an ongoing interaction,
signs that are predicted by detailed hydrodynamical simulations. We see trends
in galaxy properties, such as the truncated gaseous disks in the center
of Virgo and Coma, and truncated starformation disks in Virgo (Koopmann and
Kenney 2002) that can be reproduced in simulations of ICM interactions 
with the cold ISM in disk galaxies. These interactions should produce a 
population of early type
non star-forming disk galaxies with a range of bulge to disk ratios, as
was predicted by Dressler and Gunn (1983) and as  has been found in
Virgo (Koopmann and Kenney 2002). However other effects play a role too
in clusters. Galaxies are found that experience both stripping and 
gravitational interactions (Koopmann and Kenney 2002; Vollmer 2003) and
examples of gravitationally induced star formation have 
also been found 
(Koopmann and Kenney 2002; Rose et al. 2001; Sakai et al. 2002).
However the dominant environmental effect on cluster disk galaxies is a 
reduction of
the star formation rate, which goes hand in hand with hydrogen deficiency,
and for most galaxies this is due to ram pressure stripping 
(Koopmann and Kenney 2002).

On larger scales Solanes et al 2001 found evidence that the H~I deficiency
goes out as far as 2 R$_A$. Although this is surprisingly far, the
fact that the deficient galaxies at large distances from the cluster
tend to be on radial orbits, makes it plausible that the cause of the 
deficiency is ram pressure stripping as well. The extent of the H~I 
deficiency fits in
nicely with the results of Balogh et al. (1998), who find that the 
star formation rate as measured by the [O~II] equivalent width is depressed
in clusters out to two R$_{200}$ ($\approx$ 2 R$_A$) and more recently
the analyses of the 2dF and the SDSS survey results (Lewis et al. 2002;
Gomez et al. 2003; Nichol, this conference), which  indicate 
that star formation rates begin to drop between one and two virial radii.
These results are somewhat at odds with the
H~I imaging results where evidence is found that the groups in the
outskirts of clusters are very gas rich. Many examples of ongoing
interactions are found in these locations. In a simple scenario the
interactions would bring gas to large distances from the galaxies,
which could then easily be stripped as the galaxies fall into the denser ICM.

It is an intriguing possibility that the impact and reach of the ICM is
closely related to the dynamical state of the cluster. Cluster-(sub)cluster
merging can give rise to bulk motions, shocks and temperature structure
within the ICM. In merging clusters observational evidence has been found for 
large velocities
in the ICM (Dupke and Bregman 2001), enhanced star formation 
(Miller and Owen 2003; Miller, this conference) and distortion
of radio sources by the ICM motions (Bliton et al 1998). 
If stripping would mostly depend on the motions of the ICM and the dynamical
state of the cluster, it would more easily explain why the effects are 
seen far out into the infall region. The radial orbits 
measured for the more distant H~I deficient galaxies would then reflect
the infall direction of the most recent accretion event in the cluster.

{\bf Acknowledgments} I am grateful to the organizers of this conference for
inviting me to give this review. I thank Jose Solanes, Marc Verheijen, 
Hector Bravo-Alfaro, Dwarakanath, Jeff Kenney,  Bianca Poggianti, Raja 
Guhathakurta, Ann Zabludoff, David Schiminovich, Monica Valluri, Eric Wilcots 
and Bernd Vollmer for help with figures and many discussions on these topics. 
I thank the Kapteyn Institute,
where part of this work was done, for their hospitality. This research
was supported by an NSF grant to Columbia University and a NWO bezoekers beurs
to Groningen.


\begin{thereferences}{}


\bibitem{}
Abadi, M.~G., Moore, B., \& Bower, R.~G. 1999, \mnras 308, 947

\bibitem{}
Balogh, M.~L., Morris, S.~L., Yee, H.~K.~C., Carlberg, R.~G., \& Ellingson,
E. 1999, \apj, 527, 54

\bibitem{}
Balogh, M.~L., Schade, D., Morris, S.~L., Yee, H.~k.~C., carlberg, R.~G., \&
Ellingson, E. 1998, \apj, L78

\bibitem{}
Bliton, M., Rizza, E., Burns, J.~O., Owen, F.~N., \& Ledlow, M.~J. 1998,
\mnras, 301, 328

\bibitem{}
Bohringer, H., Briel, U.~G., Schwarz, R.~A., Voges, W., Hartner, G., \&
Trumper, J. 1994, \nat, 368, 828

\bibitem{}
Beijersbergen, M. 2003, PhD thesis, University of Groningen

\bibitem{}
Bravo-Alfaro, H., Cayatte, V., van Gorkom, J.~H., \& Balkowski, C. 2000,
\aj, 119, 580

\bibitem{}
Bravo-Alfaro, H., Cayatte, V., van Gorkom, J.~H., \& Balkowski, C. 2001,
\aa, 379, 347

\bibitem{}
Cayatte, V., van Gorkom, J.~H., Balkowski, C., \& Kotanyi, C. 1990, \aj 100,604

\bibitem{}
Cayatte, V., Kotanyi, C., Balkowski, C., \& van Gorkom, J.~H. 1994, 
\aj 107, 1003

\bibitem{}
Cowie, L.~L., \& Songaila, A. 1977, \nat, 266, 501 

\bibitem{}
Dickey, J.~M. 1997 \aj, 113, 1939

\bibitem{}
Dressler, A. 1980, \apj, 236, 351

\bibitem{}
Dressler, A. 1986, \apj, 301, 35

\bibitem{}
Dressler, A., \& Gunn, J.~E. 1983, \apj, 270, 7 

\bibitem{}
Dressler, A., Oemler, A., Couch, W.~J., Smail, I., Ellis, R.~S.,
Barger, A., Butcher, H., Poggianti, B.~M., \& Sharples, R.~M. 1997, \apj, 
490, 577

\bibitem{}
Dupke, R.~A., \& Bregman, J.~N. 2001, \apj, 562, 226

\bibitem{}
Fasano, G., Poggianti, B.~M., Couch, W.~J., Bettoni, D., Kjaergaard, P., \&
Moles, M. 2000, \apj, 542, 673

\bibitem{}
Gavazzi, G., Contursi, A., Carrasco, L., Boselli, A., Kennicutt, R.,
Scodeggio, M., \& Jaffe, W. 1995, \aa, 304, 325 

\bibitem{}
Giovanelli, R., \& Haynes, M.~P. 1983, \aj, 88, 881 

\bibitem{}
Gomez, L.~G., et al 2003, \apj, 546, 210

\bibitem{}
Gregg, M.~D., Holden, B.~P., \& West, M.~J. 2003, astro-ph/0301459

\bibitem{}
Gunn, J.~E., \& Gott, J.~R. 1972, \apj, 176, 1

\bibitem{}
Gursky, H., Kellogg, E., Murray, S., Leong, C., Tananbaum, H., \&
Giacconi, R. 1971, \apj, 167, L81

\bibitem{}
Haynes, M.~P., \& Giovanelli, R. 1984, \aj, 89, 758

\bibitem{}
Haynes, M.~P., Giovanelli, R., \& Chincarini, G.~D L. 1984, \annrev, 22, 445

\bibitem{}
Hubble. E., \& Humason, M.~L. 1931, \apj, 74, 43 

\bibitem{}
Kenney, J.~D.~P., \& Koopmann, R.~A. 1999, \aj, 117, 181

\bibitem{}
Kenney, J.~D.~P., van Gorkom, J.~H., \& Vollmer, B. 2003, \aj~submitted

\bibitem{}
Kenney, J.~D.~P., \& Young, J.~S. 1989, \apj, 344, 171

\bibitem{}
Koopmann, R.~A., \& Kenney, J.~D.~P. 2002, astro-ph/0209547

\bibitem{}
Koopmann, R.~A., \& Kenney, J.~D.~P. 1998, \apj, 497, L75 

\bibitem{}
Lewis, A. et al 2002, \mnras, 334, 673

\bibitem{}
Magri, C., Haynes, M.~P., Forman, W., Jones, C., \& Giovanelli, R. 1988, \apj
333, 136

\bibitem{}
McMahon, P.~M 1993, PhD Thesis, Columbia University

\bibitem{}
Miller, N.~A., \& Owen, F.~N. 2003, \aj, 125, 2427

\bibitem{}
Nulsen P. 1982, \mnras, 198, 1007

\bibitem{}
Oemler, A. 1974, \apj, 194, 10

\bibitem{}
Poggianti, B.~M., \& van Gorkom, J.~H. 2001, in Gas and Galaxy Evolution,
eds J.~E. Hibbard, M. Rupen, \& J.~H. van Gorkom, ASP Conf Ser 240, 599

\bibitem{}
Poggianti, B.~M., Smail, I., Dressler, A., Couch, W.~J., Barger, A.~J., 
Butcher, H., Ellis, R.~S., \& Oemler, A. 1999, \apj, 518, 576

\bibitem{}
Postman, M., \& Geller, M.~J. 1984, \apj, 281, 95

\bibitem{}
Quilis, Q., Moore, B., \& Bower, R.~G. 2000, Science, 288, 1617 

\bibitem{}
Roberts, M.~S., \& Haynes, M.~P. 1994, \annrev, 32, 115

\bibitem{}
Rose, J.~A., Gaba, A.~E., Caldwell, N., \& Chaboyer, B. 2001, \aj, 121, 793 

\bibitem{}
Sakai, S., Kennicutt, R.~C., van der Hulst, J.~M., \& Moss, C. 2002, \apj, 578,
842

\bibitem{}
Schulz, S., \& Struck, C. 2001, \mnras 328, 185

\bibitem{}
Solanes, J.~M., Manrique, A., Garcia-Gomez, C., Giovanelli, R., \& 
Haynes, M.~P. 2001, \apj, 548, 97

\bibitem{}
Spitzer, L., \& Baade, W. 1951, \apj, 113, 413

\bibitem{}
Stark, A.~A., Knapp, G.~R., Bally, J., Wilson, R.~W., Penzias, A.~A.,\& Rowe,
H.~E. 1986, \apj, 310, 660

\bibitem{}
Stevens, I.~R., Acreman, D.~M., \& Ponman, T. 1999, \mnras, 310, 663 

\bibitem{}
van den Bergh, S. 1976, \apj, 206, 883 

\bibitem{}
van Gorkom, J.~H. 1996 in the Minnesota Lectures on Extragalactic Neutral 
Hydrogen, ed. E.~D. Skillman, ASP Conf Ser 106, 293

\bibitem{}
van Gorkom et al 2003, http://www.aoc.nrao.edu/vla/html/vlahome/largeprop/

\bibitem{}
Veilleux, S., Bland-Hawthorn, J., Cecil, G., Tully, R.~B., \& Miller, S.~T.
1999, \apj, 520, 111

\bibitem{}
Verheijen, M.~A.~W., \& Zwaan, M. 2001 in Gas and Galaxy Evolution,
eds. J.~E. Hibbard, M. Rupen, \& J.~H. van Gorkom, ASP Conf Ser 240, 867

\bibitem{}
Vollmer, B. 2003, \aa, 398, 525 

\bibitem{}
Vollmer, B., Cayatte, V., Balkowski, C., \& Duschl, W.~J. 2001, \apj, 561, 708

\bibitem{}
Vollmer, B., Cayatte, V., Balkowski, C., Boselli, A., \& Duschl, W.~J. 1999,
\aa, 349, 411

\bibitem{}
Vollmer, B., Marcelin, M., Amram, P., Balkowski, C., Cayatte, V., \& Garrido,
O. 2000, \aa364, 532 

\bibitem{}
Warmels, R.~H. 1988a, \aas, 72, 19

\bibitem{}
Warmels, R.~H. 1988b, \aas, 72, 57

\bibitem{}
Warmels, R.~H. 1988c, \aas, 72, 427



\end{thereferences}
\end{document}